 \documentclass[final,5p,times,twocolumn]{elsarticle}

\usepackage{amssymb}
\usepackage{lineno}
 
\usepackage{lipsum}
\journal{Elsevier}

\begin{document}

\begin{frontmatter}

%% Title, authors and addresses

%% use the tnoteref command within \title for footnotes;
%% use the tnotetext command for the associated footnote;
%% use the fnref command within \author or \address for footnotes;
%% use the fntext command for the associated footnote;
%% use the corref command within \author for corresponding author footnotes;
%% use the cortext command for the associated footnote;
%% use the ead command for the email address,
%% and the form \ead[url] for the home page:
%%
%% \title{Title\tnoteref{label1}}
%% \tnotetext[label1]{}
%% \author{Name\corref{cor1}\fnref{label2}}
%% \ead{email address}
%% \ead[url]{home page}
%% \fntext[label2]{}
%% \cortext[cor1]{}
%% \address{Address\fnref{label3}}
%% \fntext[label3]{}

%\title{Performance of the prototype of the diamond active target of the PADME experiment at the DA$\Phi$NE BTF}
\title{Performance of the diamond active target prototype for the PADME experiment at the DA$\Phi$NE BTF}

%% use optional labels to link authors explicitly to addresses:
%% \author[label1,label2]{<author name>}
%% \address[label1]{<address>}
%% \address[label2]{<address>}
\author[INFN-LE]{R. Assiro}
\author[INFN-LE,UNI-LE]{A.P. Caricato}
\author[INFN-LE]{G. Chiodini}\ead{gabriele.chiodini@le.infn.it}
\author[INFN-LE,UNI-LE]{M. Corrado}
\author[INFN-LE,UNI-LE]{M. De Feudis}
\author[LNF]{C. Di Giulio}
\author[INFN-LE]{G. Fiore}
\author[LNF]{L. Foggetta}
\author[RM1]{E. Leonardi}
\author[INFN-LE,UNI-LE]{M.~Martino}
\author[NNL-LE,UNI-LE]{G.~Maruccio}
\author[NNL-LE,UNI-LE]{A.~G.~Monteduro}
\author[INFN-LE,UNI-LE]{F.~Oliva}
\author[INFN-LE,UNI-LE]{C.~Pinto}
\author[INFN-LE,UNI-LE]{S.~Spagnolo}\ead{stefania.spagnolo@le.infn.it}
\address[INFN-LE]{INFN - Sezione di  Lecce, Italy}
\address[UNI-LE]{Dipartimento di Matematica e Fisica ``Ennio De Giorgi" Universit\`a del Salento (LE) Italy}
\address[LNF]{Laboratori Nazionali di Frascati, 00044 Frascati (RM), Italy}
\address[RM1]{INFN sezione di Roma, Piazzale Aldo Moro 5, 00185 Rome, Italy}
\address[NNL-LE]{CNR NANOTEC - Institute of Nanotechnology (LE), Italy}

\begin{abstract}
The PADME experiment at the DA$\Phi$NE Beam-Test Facility (BTF) is designed to search for the gauge boson of a 
new $\rm U(1)$ interaction in the process e$^+$e$^-\rightarrow\gamma$+$\rm A'$, using the intense positron beam hitting a light target.  
The $\rm A'$, usually referred as dark photon, is assumed to decay into invisible particles of a secluded sector and it can be observed by searching for an anomalous peak in the spectrum of the missing mass measured in events with a single photon in the final state. 
The measurement requires the determination of the 4-momentum of the recoil photon, performed
by a homogeneous, highly segmented BGO crystals calorimeter. A significant improvement of the missing mass resolution is possible 
using an active target capable to determine the average position of the positron bunch with a resolution of less than 1 mm.
This report presents the performance of a real size $\rm (2x2 cm^2)$  PADME active target made of  a thin (50 $\mu$m) diamond sensor, with
graphitic strips produced via laser irradiation on both sides. The measurements are based on data collected in a beam test at the BTF in November 2015.
\end{abstract}

\begin{keyword}
%% keywords here, in the form: keyword \sep keyword

%% MSC codes here, in the form: \MSC code \sep code
%% or \MSC[2008] code \sep code (2000 is the default)

Polycrystalline diamond \sep Beam monitoring \sep Dark photon experiment \sep PADME 

\end{keyword}

\end{frontmatter}

%%
%% Start line numbering here if you want
%%

% \linenumbers 

%% main text
\section{Introduction}
\label{Introduction}
The compelling astrophysical evidences for the existence of dark matter still coexist with the lack of a clear experimental indication of new 
particles candidates to be constituents of this unknown form of matter \cite{citeSomeRecentReportOnDMsearches}. The LHC data have significantly pushed forward the mass constraints on dark matter candidates from standard simplified SUSY models \cite{citeATLASandCMSrecentSUSYcontraints}, thus eroding the elegance of the WIMP paradigm. Similarly, underground experiments and indirect detection experiments have steadily progressed in setting limits on interaction cross sections of dark matter of astrophysical origin with ordinary matter \cite{citeXenonEtAl} \cite{citeIndirectDMSearches}. 
In this experimental landscape, a new class of theoretical models, based on the 
hypothesis of secluded sectors of particles, where dark matter lies with no contact with the Standard Model (SM) particles, attracted an increasing attention. 
A very simple model assumes the existence of a $\rm U_d(1)$ symmetry in a new sector of particles non interacting with the SM matter. The gauge boson 
of this symmetry, a massive {\it dark} photon, $\rm A'$, can naturally mix with the Standard Model photon producing a faint indirect interaction of the SM 
matter with the dark sector. In \cite{citeReportDarkSecrorWorkshop} a comprehensive review of the existing constraints on such models from past experiments  can be found, along with a nice summary of the ongoing experimental program addressing these theoretical hypotheses. 
The PADME experiment \cite{citePADMEproposal2014} aims at searching for invisible decays of the dark photon produced in $\rm e^+e^-$ annihilation. In its baseline configuration, it will use the positron beam of 550 MeV energy produced by the LINAC of the DA$\Phi$NE \cite{citeDAFNE} complex at the Laboratori Nazionali di Frascati (LNF) of INFN, which will scatter against a $\rm 100~\mu$m thick diamond target. The experimental apparatus will be hosted in the upgraded hall of the DA$\Phi$NE Beam Test Facility (BTF) \cite{citeBTF}.  The time structure of the pulsed beam is a sequence of bunches with a controllable positron population (from single particle to about 10$^4$)  of  constant intensity in a time span of 40~ns and a repetition rate of 49~Hz.  
%The experimental apparatus consists of an electromagnetic calorimeter of cylindrical shape, for a high resolution measurement of the energy and direction of the photon \cite{citeECALnim}, a system of scintillators used as veto system for charged particles. The calorimeter is located at a distance of 3~m from the target with its axis on the incoming beam line.  Behind its central hole, a small angle calorimeter instruments the forward region of maximum flux of bremsstrahlung photons produced from the interactions of the beam with the target. The active target is located 20~cm up-stream of a dipole magnet that will sweep the beam out of the region instrumented with the calorimeters. 
The experimental apparatus consists of an electromagnetic calorimeter of cylindrical shape, for a high resolution measurement of the energy and direction of the photon \cite{citeECALnim}, with axis on the beam line, a small angle fast calorimeter behind the central hole of the main calorimeter, instrumenting the region of maximum flux of bremsstrahlung photons produced in the target, and a system of scintillators used as veto detector for charged particles. The target is located at a distance of about 3~m from the calorimeter and 20~cm up-stream of a dipole magnet that will sweep the beam out of the region instrumented with the calorimeters. The target and all veto detectors will be hosted in a vacuum chamber to minimise the interactions of the beam with the atmosphere.  
PADME will seek the signal of a dark photon as a peak on top of the smooth missing mass distribution in events with only one photon in the final state. 
The choice of a low Z material, the diamond, for the target is aimed at limiting the rate of bremsstrahlung interactions. By using the target as a detector, with the capability of reconstructing the beam 
position within $\sim 1$~mm at each bunch, the missing mass resolution improves leading to a more favourable ratio of the signal yield over the SM background. 
This report describes the performance of a full size prototype of the active target, as measured in a beam test at the BTF in November 2015.

\section{Diamond active target prototype}
\label{Diamond active target prototype}
The prototype of the active target is a sample of ``as grown" polycrystalline CVD diamond of area $\rm 2\times2~cm^2$ and $\rm 50~\mu m$ thickness, cut by a laser from a wafer produced by Applied Diamond \cite{citeAppliedDiamond}. On both surfaces 18 conductive graphitic strips separated by $\rm 150~\mu m$ have been produced with a pitch of $\rm 1~mm$. The strips are oriented in orthogonal directions on the two surfaces in order to achieve the reconstruction of the beam profile in both views. The graphitic layer is produced by irradiating the sample, mounted on a computer controlled moving support plate, with an ArF excimer laser emitting light of 193~nm wavelength in pulses of  0.9~mJ and 150~$\mu$m diameter after focusing \cite{citePaperGraphitization}.  Previous studies on these kind of electric contacts have shown that the high radiation tolerance of diamond detectors, which generally makes them interesting in high energy physics applications, is not spoiled by the graphitization procedure \cite{citeDamage}. The graphitic strips exhibit a resistance of about 2.5~$\rm k\Omega$ on the growth side (unpolished) and $\rm 5~k\Omega$ on the seed side (chemically polished). 

The sensor is mounted on a pc-board with a hole slightly smaller in size than the area on the diamond sensor.
%They are readout  by single channel front-end electronics through 15-50~cm long coaxial cables and a pc-board. 
In the beam test configuration the X strips are oriented vertically and upstream with respect to the Y strips which are oriented horizontally and are in contact with the pc-board (See Figure 1).  
%All Y strips are electrically connected to the pc-board copper traces through spots of silver paint manually deposited, while one half of X strips through wires joint by silver paint on both terminations and the remaining by aluminum wire bonding. No difference in behavior was noticed during the test between strips connected by conductive glue or wire bonding. 
The electrical connections between strips and the copper traces on the pc-board were established via spots of silver paint manually deposited for all Y strips, wires glued by silver paint on both terminations for part of the X strips, and by aluminum wire bonding for the remaining X strips. 
This procedure proved rather complex due to the poor planarity of the in-house made pc-board, the double-side detector, the small thickness of the diamond making extremely fragile the detector and the novelty of the material of the strips. 
\begin{figure}[htb]
\centerline{%
\includegraphics[width=7cm]{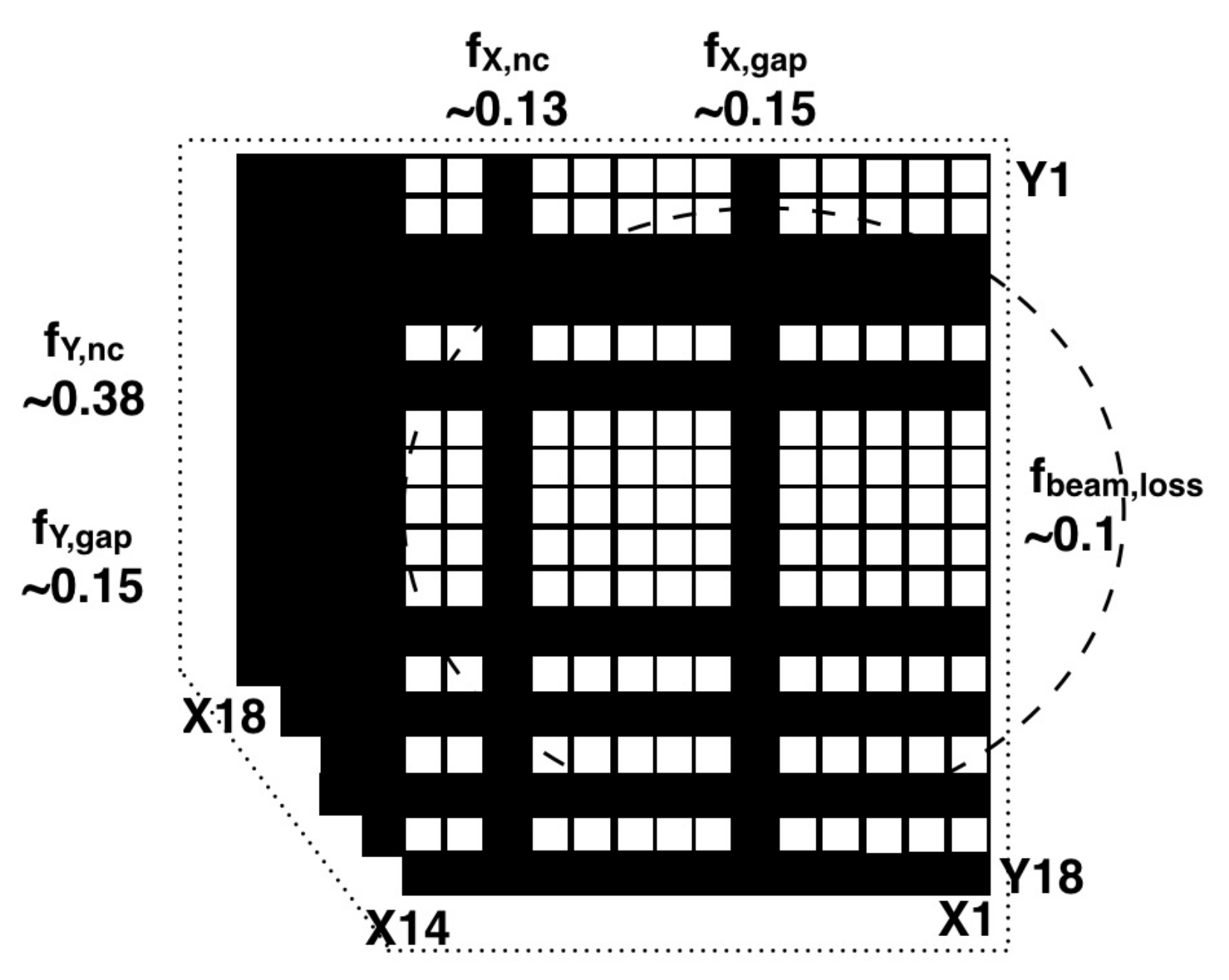}
}
\caption{Cartoon of the diamond sensor showing the biased (white color)
and the unbiased (black color) graphitic strips in the test beam.
The physical intersection between the diamond sensor ($\cdot\cdot\cdot$) and the beam (- - -)  during charge collection distance measurements  are also shown.
As explained in the text, during this measurement several sources of inefficiencies must be taken into account and their numerical values are reported.}
\label{Fig:Sensor}
\end{figure}
As a result, we suffered several failures of the contacts, leading to a reduced effective sensitive area of the detector. A sketch of the area of the detector useful for beam detection is show in Figure \ref{Fig:Sensor}. %In the beam test configuration 12 X strips , parallel to the vertical direction, on the forefront diamond surface met by the beam, and 10 Y strips turn out to be effectively connected and reading the signal induced by the beam. 
It turns out that 12 X and 10 Y strips were electrically connected, as proved a posteriori by reading out the signal induced by the beam from the corresponding electronic channels. 
The others 6 X and 8 Y strips can be considered electrically floating and the sensor volume underneath passive.
The strip numbering is also shown as seen from the view of the incoming beam. 
 In the sketch, the bottom-left corner of the square sensor is not displayed to represent a damage produced during the operations of detector assembly; the broken corner however is located in a peripheral area of the detector and therefore it does not affect its  functionality. The experience gained in the assembly of this detector prototype allowed to establish a safe 
%operation mode leading to an overall improved 
construction procedure based on manual micro-manipulators, a conductive glue dispenser, a digital camera, and vacuum pick-up tools. 

%Large size diamond detectors are not common devices in experimental high energy physics. 
Large size, above a few cm$^2$, and  small thickness,  $\leq 100~\mu$m, diamond detectors are not common devices in experimental high energy physics. The main reason is that large size mono-crystalline diamond is not affordable both from the technological and cost perspective. The only viable solution is poly-crystalline diamond grown through the CVD process, which implies a signal as lower as smaller is the size of the crystal grains. As a consequence, the production process requires the removal of a substantial fraction of the grown material from the seed side, in order to reach large size diamond grains in the final sensor. Lapping large size samples down to $\sim 100~\mu$m from an initial thickness of $\sim 1~$mm  is a very difficult and risky task. 
In our application, targeted to detect hundred of minimum ionizing particles per channel, 
%the limited quality required allows the use 
the requirements on efficiency are not too stringent and lead to the choice of a sample of CDV diamond  grown to the desired thickness, removed from the silicon substrate by chemical etching and left unpolished on the growth side. 
\begin{figure}[htb]
\centerline{%
\includegraphics[width=9cm]{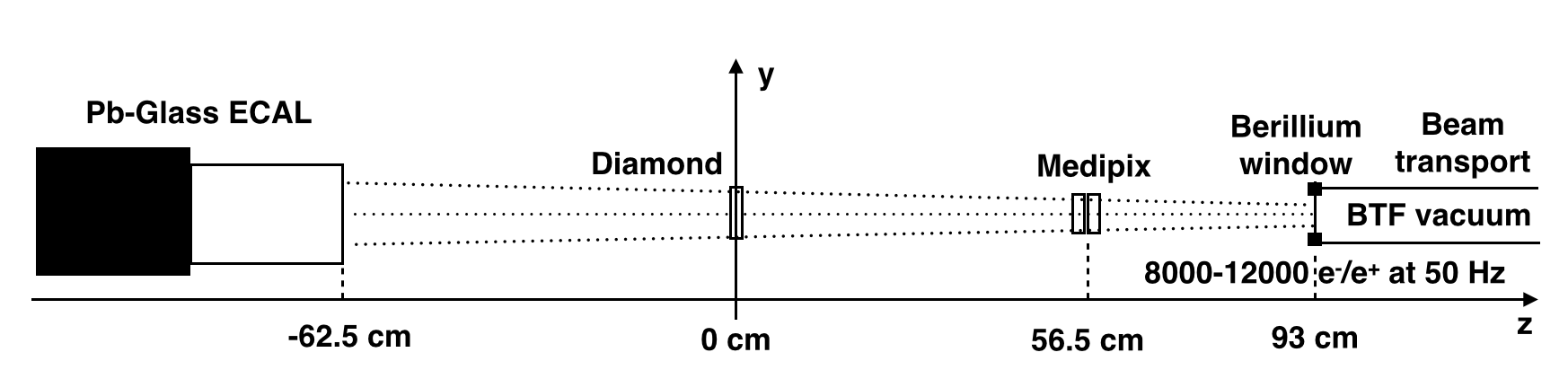}
}
\caption{Experimental setup of the beam test at the DA$\Phi$NE BTF.}
\label{Fig:ExperimentalSetup}
\end{figure}
%\subsection{Front-end electronics}
%\label{Front-end electronics}
A successful example of diamond detector, similar to the active target proposed for the PADME experiment (although based on mono-crystalline diamond) was used in the STAR experiment \cite{Ciobanu} for a fast veto system. 

\section{Experimental setup}
\label{Experimental setup}
Figure \ref{Fig:ExperimentalSetup} shows the experimental setup of the beam test at the BTF in November 2015. An aluminum support structure, mounted on a movable table, holds the diamond detector in a plane perpendicular to the beam direction; in this way translations of the detector in the two directions perpendicular to the beam can be controlled remotely. The beam leaves the vacuum pipe through a thin beryllium window about 90~cm up-stream of the PADME active target prototipe. Beyond the diamond detector, a lead glass calorimeter \cite{citeLGatBTF}, provided as a default instrumentation of the experimental area of the BTF, collects the beam and gives a measurement of the beam intensity. A hybrid silicon pixel detector of type {\tt Medipix} ~\cite{citemedipix} can be put in the beam line, in front of the diamond detector, to provide a fast diagnostic of the beam profile. It consists of a matrix of 256$\times$256 pixels of  55x55~$\rm \mu m^2$ area. 

%The strip lines on the pc-board were connected to commercial electronics optimized for diamond detectors, produced by {\tt Cividec}~\cite{citeCividec}. 
The diamond strips were readout by commercial single-channel  amplifiers optimized for diamond detectors,  produced by {\tt Cividec}~\cite{citeCividec}.
The single channel devices have a not negligible size and are connected to the pc-board through a 15-50~cm long input coaxial cable for each channel that can spoil the good signal-to-noise ratio of Charge Sensitive Amplifiers (CSA).
For comparison, half of the Y strips are readout by Voltage Amplifiers (VA) terminated by 50~$\Omega$ at the input, which typically have a worse signal-to-noise ratio with respect to CSA but are not affected by the length of the input cable.
%Voltage amplifiers (VA) terminated by 50~$\Omega$,  which typically have a worse signal-to-noise ratio but are not affected by the length of the input cable, have also been used for comparison. 
The CSA used in the beam test have a gain of 5~mV/fC, with a 100 MHz bandwidth, 2 ns rise-time, 7 ns pulse width and 750~$e$  equivalent input noise. The VA have a gain of 100, 2 GHz bandwidth and 25~$\mu$V equivalent input noise when terminated with 50~$\Omega$ at the input.
%A successful example of diamond detector, similar to the active target proposed for the PADME experiment, was used in the STAR experiment \cite{Ciobanu} for a fast veto system. 
The data acquisition was based on multichannel digitizers  CAEN V1742 ~\cite{citedigitizerCAEN} sampling the input signal at 1, 2.5 or 5~GS/s with 12 bit ADC resolution over a dynamic range of $\pm$0.5~V+offset(-0.5~V) and a buffer depth of 1024 bins. The same readout electronics was reading the signal from the lead-glass calorimeter after attenuation. Most of the data were collected with a sampling rate of 1~GS/s and a beam of interleaved spills of electrons and positrons  with  an average energy of 450 MeV.  The typical average multiplicity of particles in a bunch was 10000 with a higher multiplicity ({about 20\%) in case of electron spills. During most of the data taking the active target was operated with a bias voltage of 150~V, delivered internally by the amplifier inputs connected to the X strips.  The DAQ system, based on a prototype version of the PADME data acquisition and online monitoring software \cite{citeLeonardiCHEP} was recording the entire digital waveforms for each channel (36 signals from the diamond strips and one signal from the calorimeter) at the rate of the beam clock. 
\begin{figure}[htb]
\centerline{%
\includegraphics[width=9cm]{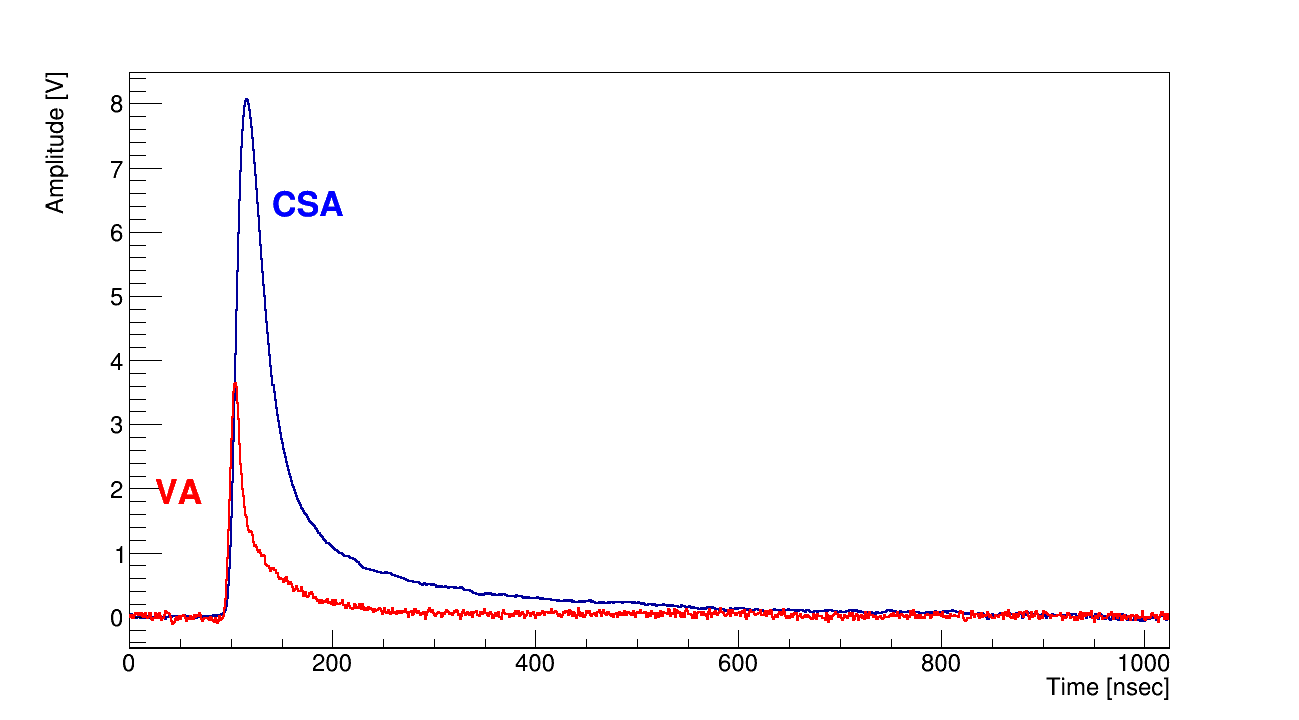}
}
\caption{Sum of digitized output signals of 300 events from the fast charge amplifier (CSA) connected to the X9 strip  (dotted line) and from the fast voltage amplifier (VA) connected to the Y11 strip (continuous line). The detector was operated at a bias voltage of 150~V.\label{Fig:Signals}}
\end{figure}

%\begin{figure}[t]
%centerline{%
%\includegraphics[width=9cm]{./signalsRF_CSA-eps-converted-to.pdf}
%}
%\caption{Sum of digitized output signals of 300 events from the fast charge amplifier connected to the X9 strip  (dotted line) and from the fast voltage amplifier
%connected to the Y11 strip (continuous line). The detector was operated at a bias voltage of 150~V. 
%\label{Fig:Signals}
%\end{figure}

Figure \ref{Fig:Signals} shows, superimposed, the digitized output signals from two strips, one connected to a charge amplifier and the other connected to a  voltage amplifier. In order to highlight the features of the two kind of signals by suppressing the fluctuations due to the noise of the electronic readout chain, the waveforms of 300 consecutive events have been summed. The peaking time of the  VA output is about equal to the duration of the particle bunch, as expected, while the CSA peaking time is about two times longer. 
%The rise time and the fall time of the two signals are pretty much similar, likely both dominated by the time constant of the strips. 
The not negligible ohmic resistance of the strips affects the shape of the output signal smoothing the response of the amplifiers. The response of the same electronics on a similar sensor with metallic strips is reported in \cite{citeChiodiniProc2016}. 

For each event, the waveform recorded for each strip of the diamond detector has been processed to extract the charge collected due to the passage of the beam. 
This is done by integrating the signal in the range 80~ns $-$ 500~ns and applying a multiplicative calibration constant previously measured for each amplifier. The calibration procedure was made on bench by injecting a known charge via a rectangular pulse lasting 10 ns, emulating the time structure of a particle bunch of the beam.
The charge measured by the strips in each plane as a function of the strip position provides a reconstruction of the beam, bunch per bunch. Examples of beam profiles as seen by the PADME active target are shown in Figure \ref{Fig:BeamProfilesX} for the X (horizontal) view and Figure \ref{Fig:BeamProfilesY} for the Y (vertical) view. To suppress fluctuations, the profiles obtained in 300 consecutive events recorded under stable beam conditions, are summed. In the two plots the blue histogram represents real measurements, while the red bins correspond to disconnected strips and the charge is estimated via a linear interpolation of the charge measured in the neighbouring strips. 
\begin{figure}[htb]
\centerline{%
\includegraphics[width=8cm]{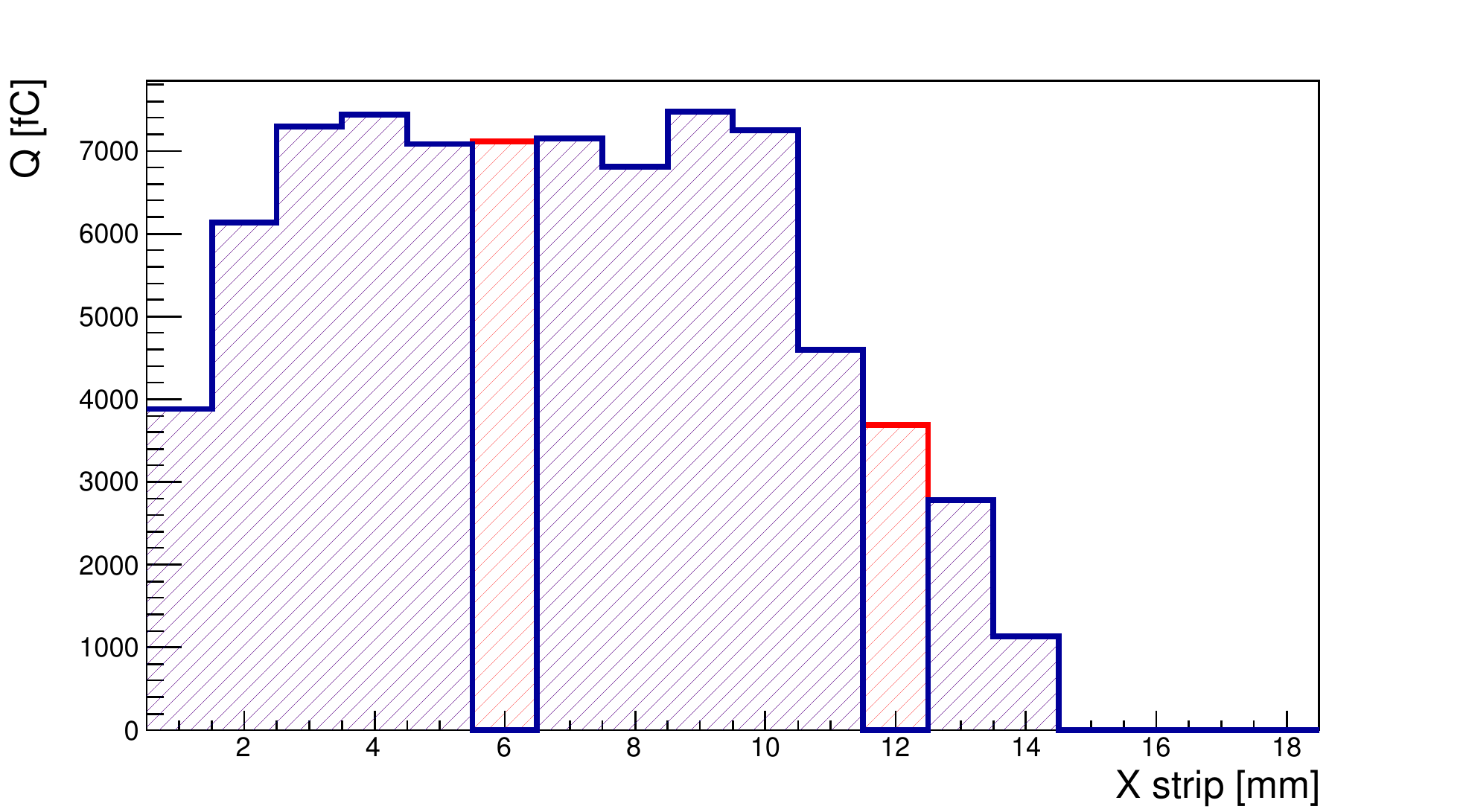}
}
\caption{Sum of the beam profile in the horizontal direction as measured by the diamond detector in 300 events. The entries for strips 6 and 12 are obtained by a linear interpolation of the charge measured in the neighbouring strips. The detector was operated at a bias voltage of 150~V. All strips are readout by charge amplifiers.}
\label{Fig:BeamProfilesX}
\end{figure}

\begin{figure}[htb]
\centerline{%
\includegraphics[width=8cm]{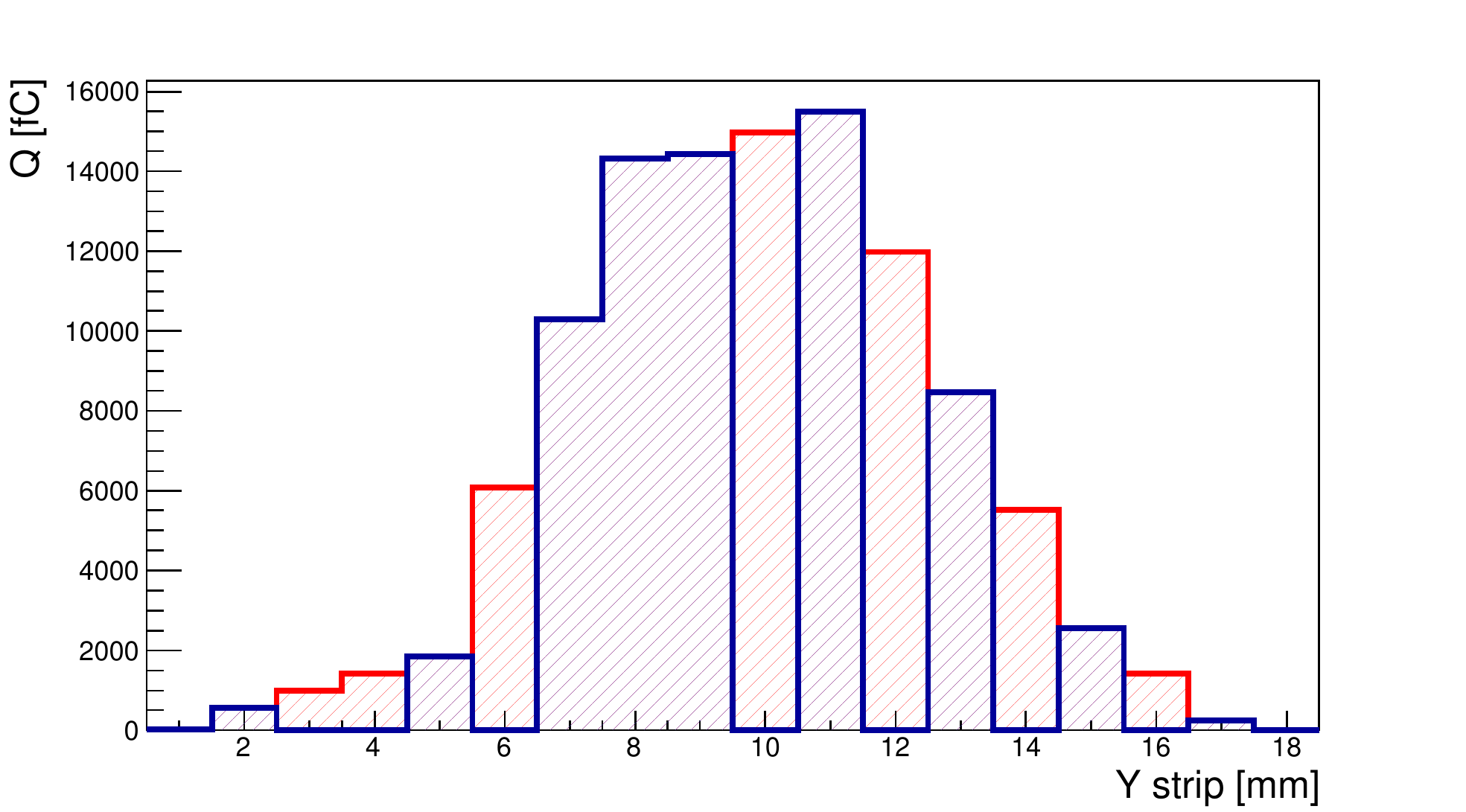}
}
\caption{Sum of the beam profile in the vertical direction as measured by the diamond detector in 300 events. The entries for strips 3, 4, 6, 10, 12, 14 and 16 are obtained by a linear interpolation of the charge measured in the neighbouring strips. The detector was operated at a bias voltage of 150~V. Strips 2, 7, 8, 11 and 15  were readout by voltage amplifiers, while the remaining strip by charge amplifiers.}
\label{Fig:BeamProfilesY}
\end{figure}

Figure \ref{Fig:ChargeTrend} shows the sum of the charge measured in the event by all X (middle plot) and Y (bottom plot) strips as a function of the event number; in the top plot, the corresponding particle multiplicity measured by the calorimeter is displayed for reference. The alternating sequences of positron and electron spills can be distinguished thanks to the lower average intensity of the positron beam. 
The plot indicates that the intensity of the beam is not stable; in addition it suggests that the positron and electron beam have different profiles and central position, therefore the target intercept a smaller fraction of the beam profile in the case of the positron beam that in the case of the electron beam. This was confirmed during data taking by inspecting the beam profiles recorded by the pixel detector, which showed also a small drift of the beam center with time.
\begin{figure}[htb]
\centerline{%
\includegraphics[width=8cm]{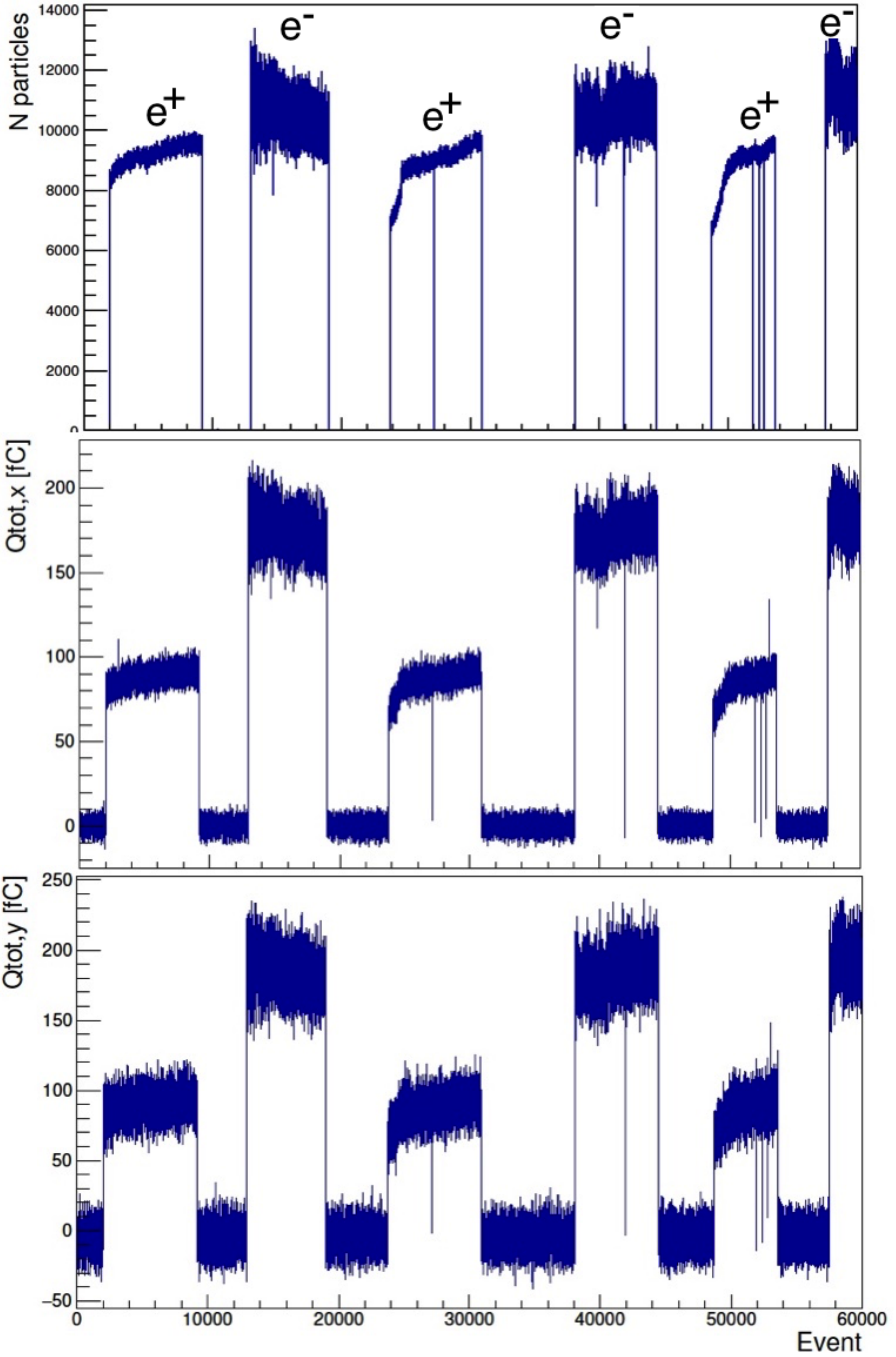}
}
\caption{Evolution with time of the electron and positron bunch multiplicity as measured by the calorimeter (top), by the diamond detector from the instrumented X plane (middle), by the 
diamond detector from the instrumented Y plane (bottom).
\label{Fig:ChargeTrend}}
\end{figure}

\section{Results}
\label{Results}
The data collected in the beam test have been used to demonstrate the efficiency of the polycrystalline diamond with graphitic strips as monitor detector for intensity and position of the electron/positron beam of the DA$\Phi$NE complex. A quantitative measurement of the detector response is provided by the measurement of the charge collection distance of the device, described in the following, that allows to understand the signal dependence on the beam intensity. Another important parameter is the spatial resolution on the beam centroid, which impacts on the resolution in invisible mass. Beyond the scope of the PADME experiment, the time resolution of the device is interesting on its own due to the very fast response of diamond detectors and the increasing demand of detectors with high time resolution in high energy physics. 

\subsection{Time resolution}
\label{Time resolution}
The shape as a function of time of the output signals from the strips has been fit to extract the time stamp $t^\star_0$ of the moment the signal sets in. 
The analytic expression of the function $V(t)$ used in the fit is 
\begin{eqnarray}
\begin{tabular}{ll}
$V_0$ & $t<t_0$ \\
$V_0+V_1\frac{t-t_0}{t_1}e^{-\frac{t-t_0}{t_1}}P_3(t)$ & $t>t_0$
\end{tabular}
\end{eqnarray}
%\begin{equation}
%\begin{sistema}
%V_0 \qquad \qquad \qquad \qquad \quad \ \  \quad per \quad t<t_0 \\
%V_0+V_1\frac{t-t_0}{t_1}e^{-\frac{t-t_0}{t_1}}P_3(t) \qquad per \quad t>t_0
%\end{sistema}
%\end{equation}
where $V_0$ is the baseline value and $P_3(t)$ is a third degree polynomial with the constant term fixed to one, and two free parameters, $V_1$, $t_0$ and $t_1$ are determined from the fit. After convergence, the fit is restricted to the front of the signal, i.e. $t<t_{max}$ where $V(t_{max}) = 0.75 \times V_{max}$, in order to avoid any bias from mismodeling of the tail of the signal shape; $t^\star_0$ is defined as the best fit value of $t_0$ in the second fit.  
In order to minimise the contribution to the dispersion of the values of $t^\star_0$ due to fluctuations related to the beam, the time resolution has been estimated from the distribution of the difference in $t^\star_0$ between two adjacent strips which were approximately seeing a similar fraction of the beam profile in a run collected in rather stable conditions and setting the sampling rate of the digitizer to 5~GS/s. The distribution is shown on Figure \ref{Fig:TemporalResolution}. From the width of the distribution a time resolution of $\sim$2~ns per strip can be inferred.

\begin{figure}[htb]
\centerline{%
\includegraphics[width=8cm]{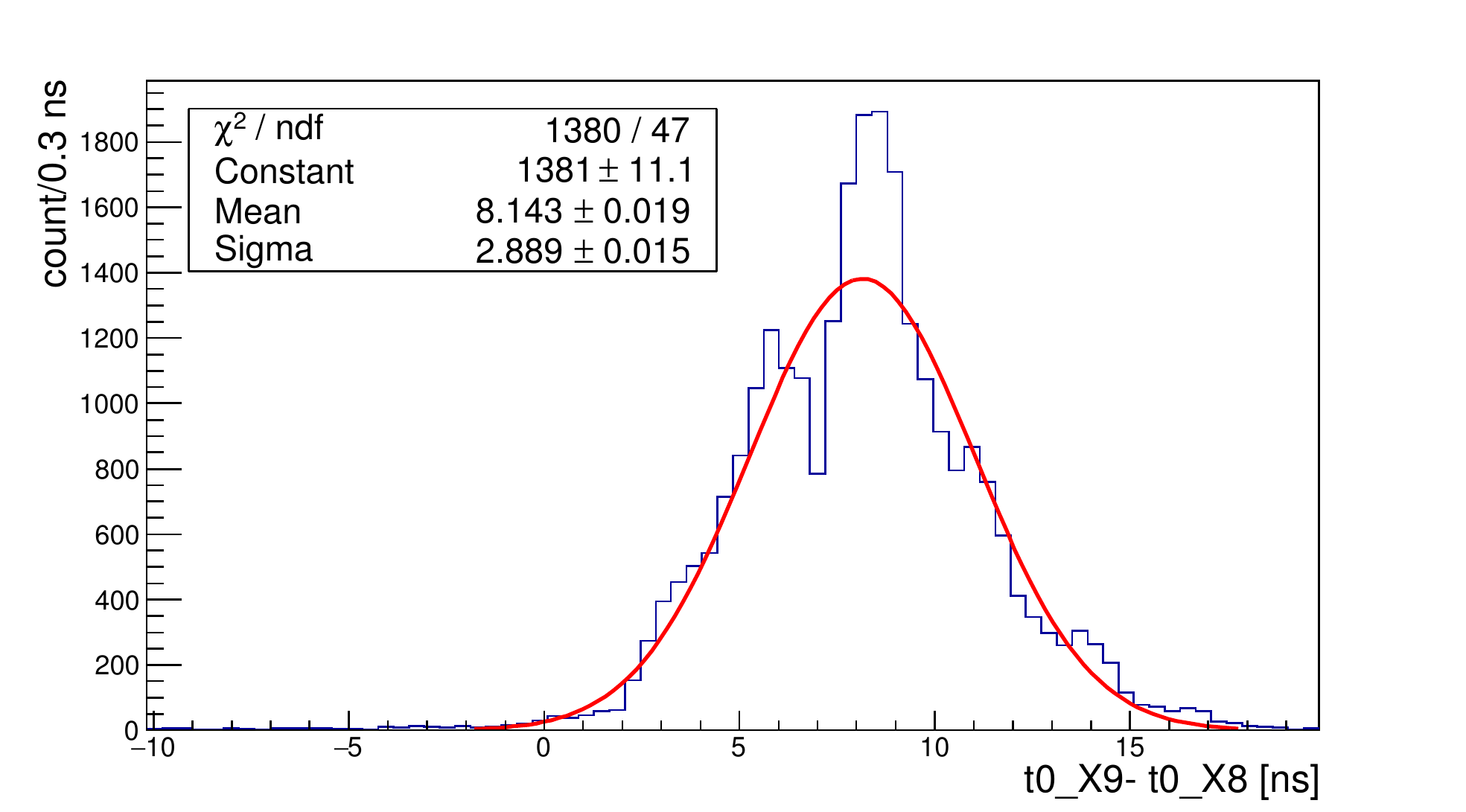}
}
\caption{Distribution of the signal arrival time difference for strips X8 and X9. The detector was operated at a bias voltage of 150~V.}
\label{Fig:TemporalResolution}
\end{figure}

\subsection{Charge collection distance}
\label{Charge collection distance}
In polycrystalline diamond various kinds of defects, acting as charge trapping centres, prevent the complete collection of ionization charges to the electrodes. The parameter typically used to characterise the quality of the material as radiation detector is the charge collection distance (CCD) defined as the ratio of the induced charge $Q$ (integral of the  current signal) over the ionization charge, $N_{ion} |e|$,  multiplied by the detector thickness $L$
$$
{\rm CCD} = L \cdot \frac{Q}{N_{ion} |e|}. 
$$
Since the mean free path of charge carriers increases with their velocity the CCD depends on the value of the electric field within the diamond film. The beam test data allow to measure the CCD of the CVD polycrystalline diamond of the detector using the following relation: 
\begin{equation}
{\rm CCD~ [\mu m]} = \frac{Q {\rm [ fC ] \cdot 6250 [e^- / fC ]}} {  {\rm 36 [e^-/\mu m] \cdot N_{active} } }
\label{eqCCD}
\end{equation}
where the total charge $Q$ is the sum of the induced charges on all X strips (or Y strips). 
In equation \ref{eqCCD} the average number of electron-ion pairs produced by a m.i.p. traversing diamond is assumed to be equal to $\rm  36/\mu m$.  The number of particles of the beam hitting the active area of the detector, $\rm N_{active}$, is smaller than the total beam intensity due to geometrical losses (10\% of the beam profile is outside the area of the active target, as can be seen from the beam profile show in Figure \ref{Fig:BeamProfilesX}), inactive area of the detector (15\% of each surface is not biased due to the inter-strip dead gaps) and dead (not well connected) strips (the fraction of dead strip area intersecting the beam is 13\% in the plane measuring X and 38\% in the plane measuring Y). 
%Since in the run used for the CCD measurement the beam intensity, measured from the calorimeter, was 10$^4$,  the effective number of m.i.p. hitting the detector is $\rm N_{active}=3500$. 
%A dedicated high statistics run has been used to estimate 
The value of the total charge $Q$ in equation \ref{eqCCD} has been evaluated as the mean of a gaussian fit to the distribution of $Q$ in a dedicated run, where the detector was operated at a bias voltage of 150~V. The beam intensity, measured by the calorimeter, was 10$^4$,  hence the effective number of m.i.p. hitting the detector was $\rm N_{active}=3500$. 
The resulting measurement  $$\rm CCD = 11.35 \pm 0.03 (stat.) \pm 0.45 (cal.) \pm 0.05 (other~syst.)~\mu m$$ is affected by the dominant systematic uncertainty on the calibration  (4\%). 
The difference between the measurement from the X and the Y plane (1\%) is used as an estimate of other residual systematic effects. 
The CCD value obtained is satisfactory, given the quality of the entire process for the production of the diamond sample used as sensor and the resulting low cost of the device. 
The same procedure was applied to analyse data of a run where the positive bias voltage was varied from 175~V to 50~V and of another run where the negative bias was increased from $-50$~V to $-175$~V. The trend of the CCD as a function of the voltage difference between the two surfaces, is shown in Figure \ref{Fig:CCD}. The measurements are in agreement within uncertainties with the CCD measured by the manufacturer on similar samples after pumping with beta source. Our measurements are however systematically lower as expected due to the missing pumping treatment. At low field values, some discrepancy can be observed between the performance of the detector under positive and negative bias, presumably indicative of some polarization effect. 

%%%%%%% NOTA: l
%{\it La misura puntuale e' fatta a V=150 V con una f$_{active}$ diversa da quella delle condizioni di run della figura ... commentare.}
\begin{figure}[htb]
\centerline{%
\includegraphics[width=8cm]{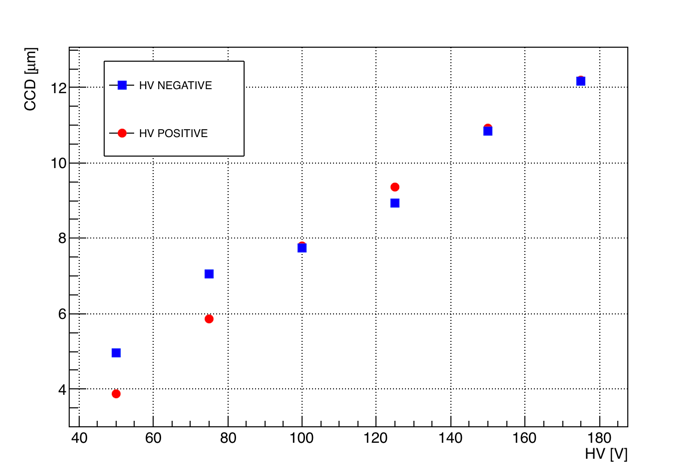}
}
\caption{Measured average charge collection distance for different positive and negative high voltage values. The error bars representing the statistical errors do not exceed the size of the markers.}
\label{Fig:CCD}
\end{figure}

With the front-end electronics used in this beam test and the signal transport chain used for the prototype of the PADME active target (including coaxial cables of length between 15 and 50~cm from the pc-board holding the detector to the amplifiers) the typical equivalent input noise, estimated by applying the same reconstruction procedure as in the case of a distinct signal, is 
$$
{\rm Q^{noise}_{CSA} = 0.8~fC,\ \ \  Q^{noise}_{VA} = 3.6~fC}. 
$$
These values, along with the measurement of the CCD presented earlier, guarantees a satisfactory signal to noise ratio ($S/N$) even for a rather low intensity of the bunch in the beam: 
\begin{eqnarray}
\begin{tabular}{ll}
$S ({\rm 1000 m.i.p.})/N \simeq 80$ & for CSA \\
$S ({\rm 1000 m.i.p.})/N \simeq 18$ & for VA \\
\end{tabular}
\nonumber
\end{eqnarray}
After this beam test, a different kind of front-end electronics has been used to read the signal from the detector exposed to a similar configuration of the BTF beam with satisfactory results \cite{citeChiodiniProc2016}.

\subsection{Spatial resolution}
\label{Spatial resolution}
From the beam profiles the position of beam centroid in X and Y can be estimated by the average of the strip positions, in the two planes, weighted with the measured (or estimated in case of disconnected strips) charge.  The estimate of the resolution on the beam centroid position is obtained by the sigma of a gaussian fit to the distribution of the reconstructed beam centroid in 300 events recorded under stable beam conditions, with $10^4$ particles per bunch and a X-Y beam spot of $\sim 2~mm x 3~mm$. Figure \ref{Fig:SpatialResolution} shows the distribution of the X coordinate of the beam centroid. The width of the distributions obtained for the X and Y coordinate are $0.20 \pm 0.01$~mm and $0.32\pm 0.02$~mm respectively, well below the value ($\simeq 1~mm$) of spatial resolution on the beam centroid  required by the PADME experiment. However, it's worth mentioning that the dispersion of the values of the beam centroid shown in Figure \ref{Fig:SpatialResolution} arises both from the detector resolution and from the intrinsic instability of the beam position. The latter can be estimated from the dispersion of the beam centroid position measured with {\tt Medipix}, where the contribution of the resolution of the pixel detector is negligeable. This measurement was derived from dedicated data collected previously for beam diagnostic purposes and in different beam multiplicity conditions; after scaling to the proper number of particle per bunch, the contribution from the stability of the beam in the width of the distribution of Figure \ref{Fig:SpatialResolution} is estimated of the order of 150~$\mu$m. 
\begin{figure}[htb]
\centerline{%
\includegraphics[width=9cm]{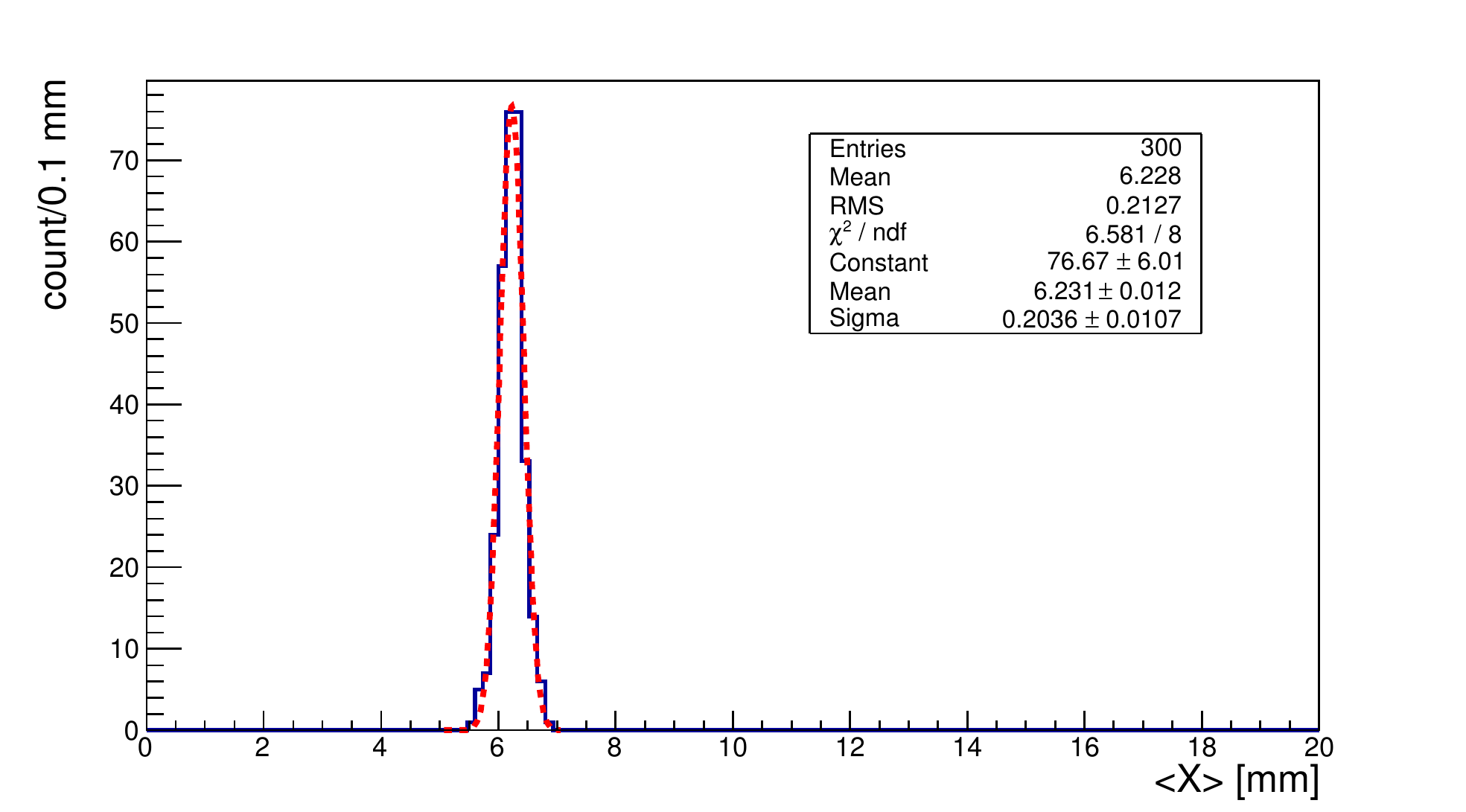}
}
\caption{Distribution of the average  horizontal beam position as measured by the charge weighting algorithm as explained in the text. Data are collected at HV=150~V.}
\label{Fig:SpatialResolution}
\end{figure}

During a run the vertical position of the movable table holding the active target prototype was changed in steps of 1~mm after about 10~s of data taking. 
The trend of the beam charge centroid in Y measured event by event by the detector is shown by the blue histogram in Figure \ref{fig:Linearity}. A selection of 280 events for each step (shown in red, with errors corresponding to the RMS of the beam centroid around the average value in each region) are used in the fit. The analytic function used in the fit is the sum of a linear function and a stair of constant size steps:  
$$
	f (x) = p0 + p1 \cdot ( x - x0 ) + j \cdot \Delta y 
$$
where $j$ is the step index, ranging from 1 to 10. 
A fit with $\Delta y$ constrained to 1~mm and both the constant term and the common slope floating gives a reasonable $\chi^2/$ndf = 1.4. The best fit value of the common slope is compatible within two standard deviations with the slopes, $p1_i$ (with $i$=1,...,10), obtained from independent fits of each region. On the other hand, the best fit of $p1$ appears significantly lower than the average of the slopes $p1_i$. This is an indication  that the analytic expression used for the fit is not able to model some systematic effects. 
In fact, several important systematic errors are expected as a result of non linear or non constant drift of the beam centroid with time, asymmetric and not fully contained beam cross section, missing measurement for the unconnected strips, and residual mis-calibrations of the amplifiers.  However, within the limited control of the systematic uncertainties affecting the measurement, the successful fit with $\Delta y$ constrained to 1~mm demonstrates the linearity of the beam centroid measurement with respect to changes of the beam-target relative position. In the experiment the beam profile will be much narrower and regular in shape and the related systematics are expected to be under control.

%A fit with all three parameters left free (yellow line in the plot) gives a best fit value of $\Delta y = 0.757 \pm 0.036$~mm. However, from Figure \ref{fig:Linearity}, it can be observed that the beam position does not appear to be very stable in the regions of constant position of the detector. If $\Delta y$ is constrained to the value of 1~mm, the fit gives a common slope value lower than the best fit value of the slope in each individual step, but still compatible with all of them within two standard deviations. 
%A final fit, in cyan in  Figure \ref{fig:Linearity}, with the slope constrained to the average of the slopes measured in each individual step provides an estimate of $\Delta y = 0.8256 \pm 0.0027$~mm. All fit options have comparably good $\chi^2/ndf$. The values of the dispersions of the measured Y beam centroid with respect to the two fitting functions, with fixed or free $\Delta y$, differ by less than 1\%, and they are about 2\% higher than the dispersion with respect to the average in each step. Therefore, within the limited control of the systematic uncertainties affecting the measurement, the successful fit with $\Delta y$ constrained to 1~mm demonstrates the good linearity of the reconstructed beam centroid with respect to beam displacements. 

\begin{figure}[htb]
\centerline{%
\begin{tabular}{c}
\includegraphics[width=8cm]{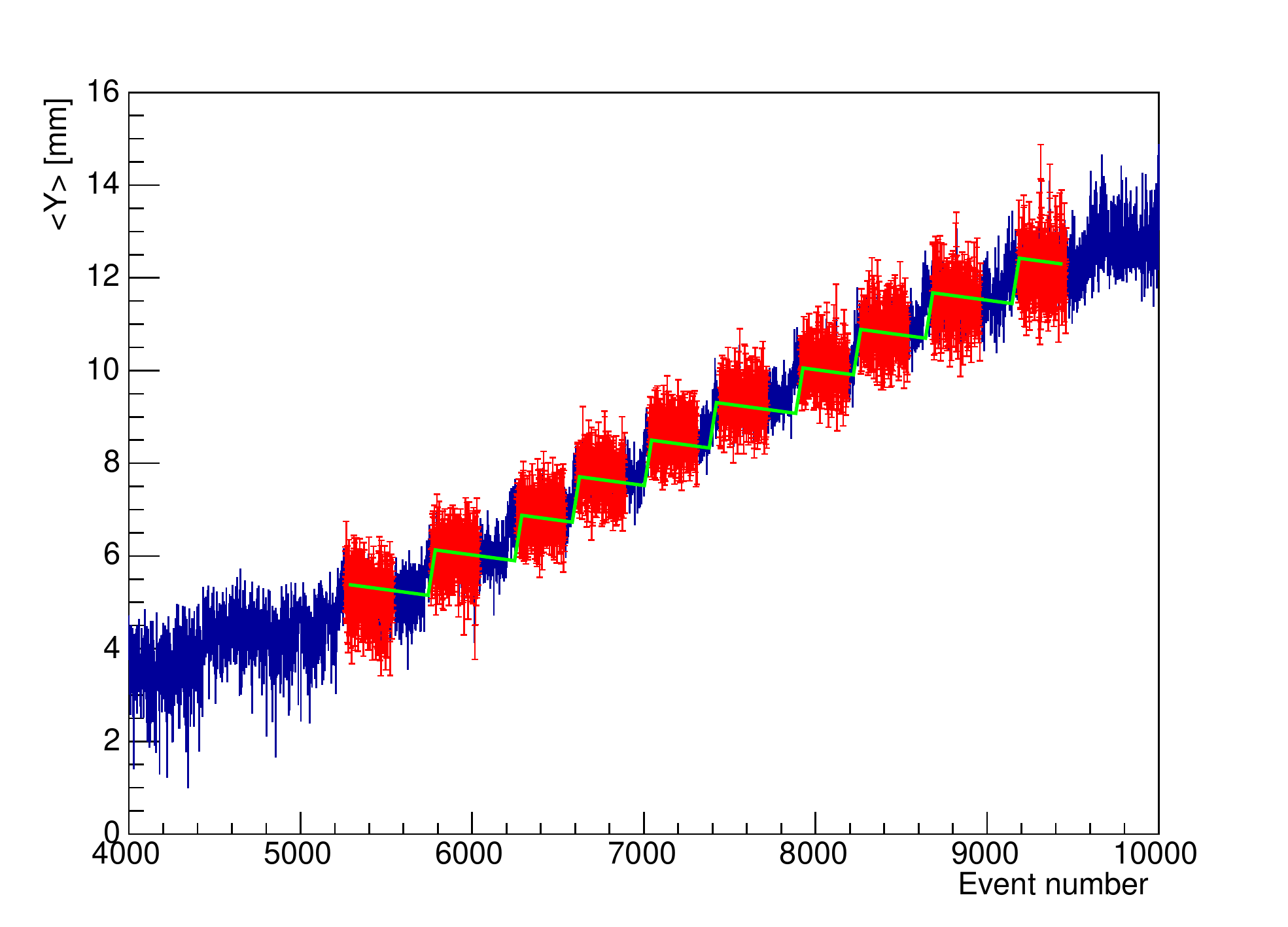} \\
\includegraphics[width=8cm]{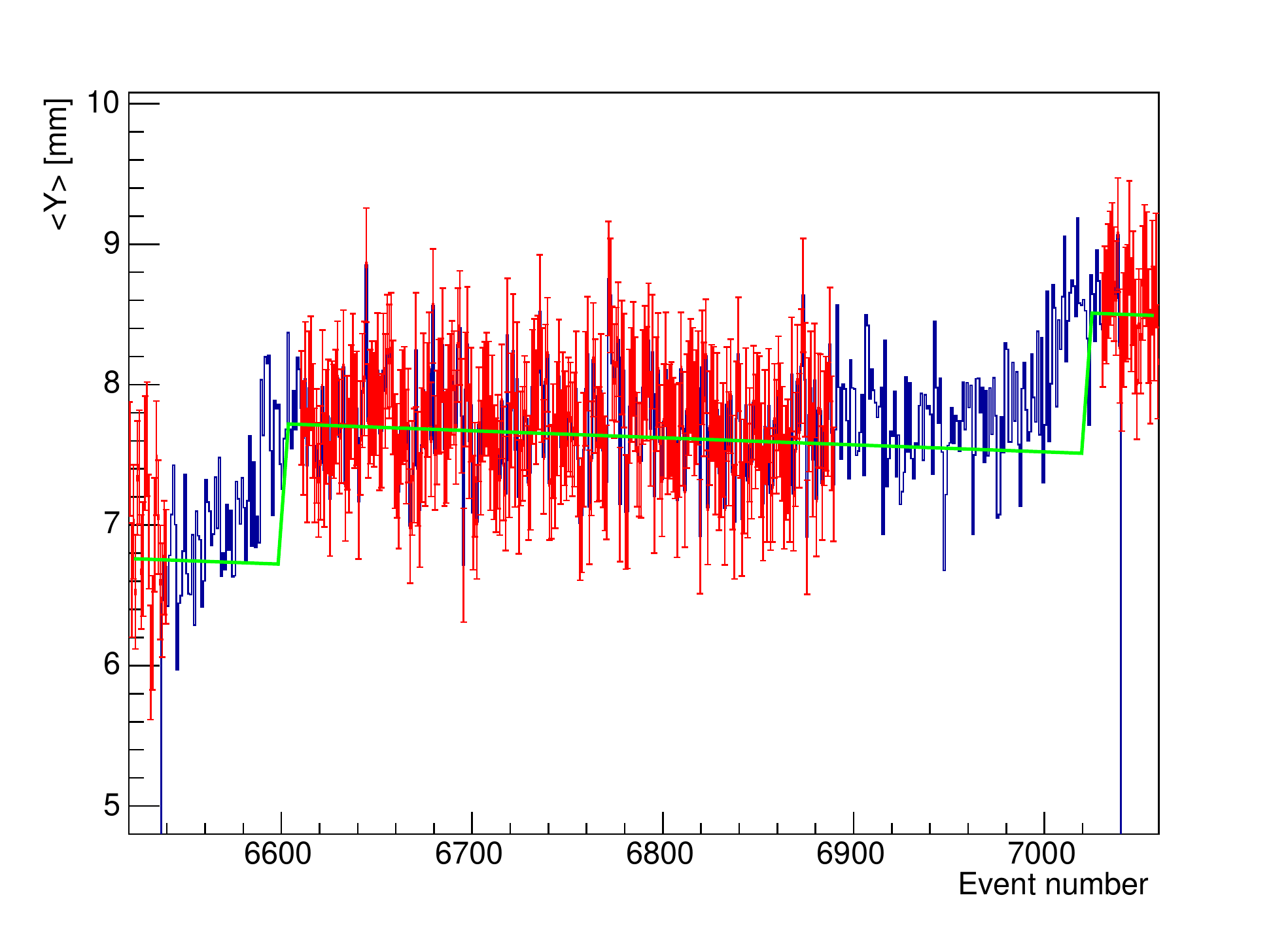} \\
\label{fig:Linearity}
\end{tabular}
}
\caption{Y coordinate of the beam charge centroid as a function of the event number in a run where the vertical position of the active target has been moved in steps of 1~mm; the events in red are used in several fit (continuous lines in yellow, blue, green and cyan) with a step function where each step is a linear function. In the bottom plot a zoom of a central step is shown. Data are collected at HV=150~V.}
\label{Fig:Linearity}
\end{figure}

\section{Conclusions}
A large size and thin detector, $20\times 20\times 0.05~mm^3$, built from an ``as grown" polycrystalline diamond sample, with laser made graphitic strips on both sides, has been tested with the electron-positron beam of 
%An innovative detector, consisting of a large area thin polycrystalline diamond sample $\rm 20 \times 20 \times 0.05~mm^3$ with graphitic strips on both sides, has been tested with the positron beam of 
the Beam Test Facility of the DA$\Phi$NE complex at LNF. The detector has been readout with amplification electronics optimised for fast signals of low amplitude. 
%A good detection efficiency is observed, although the measured charge collection distance is of the order of $\rm 10~\mu m$ for an electric field of $\rm 3~V/\mu m$, rather low as expected based on the quality of the CVD production process for large area devices.  
A good and uniform detection efficiency is observed. The measured charge collection distance is of the order of 10~$\mu$m for an electric field of 3~V/$\mu$m without pumping, as expected for such a 
thin and unlapped CVD  diamond sensor.
A time resolution of $\rm 2~ns$ and a space resolution of about $\rm 200~\mu m$ on the charge centroid  are measured for a high intensity beam (about $10^4~e^\pm$ per bunch). The capability of the detector to follow variations of the beam position with time also demonstrated. The performance of the detector measured in the beam test meets all requirements of the active target of the PADME experiment and more generally shows that this kind of detector could be used as an ``in line" radiation monitor for moderate and high intensity beams. The results suggest that less demanding and expensive front end electronics, as available in standard  multi-channel  readout chips for silicon detectors, with a bigger integration time and similar input noise, can still be suitable for the purposes of the experiment. 

\section{Acknowledgments}
We warmly thank the BTF and LINAC teams, for the excellent quality of the beam. The authors are also grateful to E. Raggi, P. Valente and the whole PADME team for general support and guidance during the beam test. 
%E. Leonardi for having integrated the active target in the Trigger/DAQ system in operation at the beam test and for his continuous support to the data handling and analysis framework.  
This work is partly supported by the project PGR-226 of the Italian Ministry of Foreign Affairs and International Cooperation (MAECI), CUP I86D16000060005.


\begin{thebibliography}{10}

%1
 \bibitem{citeSomeRecentReportOnDMsearches} G. Bertone, D. Hooper, \emph{ A History of Dark Matter}, arXiv:1605.04909 [astro-ph.CO] (2016)
 
 %2
 \bibitem{citeATLASandCMSrecentSUSYcontraints} W. Adam {\it on behalf of the ATLAS and CMS Collaborations}, \emph{Searches for Supersymmetry}, Proceedings of the 38th International Conference on High Energy Physics,  PoS(ICHEP2016) 017
 
 %3
 \bibitem{citeXenonEtAl} Jianglai Liu,	Xun Chen	and Xiangdong Ji, \emph{Current status of direct dark matter detection experiments}, Nature Physics 13, 212?216 (2017)
 
 %4
 \bibitem{citeIndirectDMSearches} J. M. Gaskin, \emph{A review of indirect searches for particle dark matter}, Contemporary Physics 57,4 (2016) 496
  
  %5
 \bibitem{citeReportDarkSecrorWorkshop} Jim Alexander {\it et al}, \emph{Dark Sectors 2016 Workshop: Community Report}, 	arXiv:1608.08632 [hep-ph] (2016)
 
 %6
 \bibitem{citePADMEproposal2014} M. Raggi and V. Kozhuharov, \emph{Proposal to Search for a Dark Photon in Positron on Target Collisions at DA$\Phi$NE Linac}, Adv. High Energy Phys. 2014, 959802 (2014)
 
 %7
 \bibitem{citeDAFNE} A. Ghigo, G. Mazzitelli, F. Sannibale, P. Valente and G. Vignola, \emph{Commissioning of the DA$\Phi$NE beam test facility}, Nucl. Instrum. Meth. A 515, 524 (2003)
 
 %8
 \bibitem{citeBTF} P. Valente et al., \emph{Linear Accelerator Test Facility at LNF Conceptual Design Report}, arXiv:1603.05651 [physics.acc-ph]
 
 %9
 \bibitem{citeECALnim} M. Raggi {\it et al.}, \emph{Performance of the PADME calorimeter prototype at the DA$\Phi$NE BTF}, Nucl. Instrum. and Methods A862, 31 (2017)
 
 %10
 \bibitem{citeAppliedDiamond} http://usapplieddiamond.com

%11
\bibitem{citePaperGraphitization}
M. De Feudis {\it et al.}, \emph{Diamond graphitization by laser-writing for all-carbon
detector applications},  Diamond \& Related Materials 75, 25 (2017)

%12
\bibitem{citeDamage} 
G.Chiodini {\it et al.}, \emph{Radiation damage of polycristalline diamond exposed to 62 MeV protons}, Nucl. Inst. Methods A {\bf 730}, 152-154 (2013)

%13
\bibitem{Ciobanu}
M. Ciobanu {\it et al.}, \emph{In-Beam Diamond Start Detectors}, IEEE Trans. Nucl. Sci., Vol.58, No. 4, 2073 (2011)

%14
%\bibitem{Spieler}
%H. Spieler,  \emph{Fast Timing Methods for Semiconductor Detectors}, IEEE Trans. Nucl. Sci. NS-29/3 (1982) 1142

 %15
\bibitem{citeLGatBTF}
https://wiki.infn.it/strutture/lnf/da/btf/home
 
 %16
\bibitem{citemedipix}
http://medipix.web.cern.ch/medipix
 
%17
\bibitem{citeCividec}
http://www.cividec.at 
 
 %18
 \bibitem{citedigitizerCAEN} CAEN Mod. 1742, Technical Information Manual,
rev6 06 February 2016,  http://www.caentechnologies.com/
 
  %19
 \bibitem{citeLeonardiCHEP}
 E. Leonardi {\it et al.}, \emph{Development and test of a DRS4-based DAQ system for the PADME experiment at the DAFNE BTF} to appear in Proceedings of CHEP 2016. 

%21
\bibitem{citeChiodiniProc2016} 
G.~Chiodini and the Active Target PADME Collaboration, \emph{A diamond active target for the PADME experiment},  JINST 12, C02036 (2017).


% \bibitem{Cartiglia}
% N. Cartiglia et al. \emph{Performance of ultra-fast silicon detectors}, JINST {\bf 9} C02001 (2014).

%20
\bibitem{Foggetta}
  L.~G.~Foggetta, B.~Buonomo and P.~Valente.  \emph{Beam Optimization of the DA$\Phi$NE Beam Test Facility}.
 Proceedings of 6-th International Particle Accelerator Conference (IPAC 2015). Richmond (Virginia, USA), May 3-8, (2015).


\end{thebibliography}
\end{document}